# Plasma Instability and Amplified Mode Switching Effect in THz Field Effect Transistors with Grating Gate

G. R. Aizin[1,*], J. Mikalopas[1], and M. Shur[2,†]

[1]*Kingsborough College, The City university of New York, Brooklyn, New York 11235*

[2]*Rensselaer Polytechnic Institute, Troy, New York 12180*

**Abstract**

We developed a theory of collective plasma oscillations in a dc current-biased field effect transistor with interdigitated dual grating gate and demonstrated a new mechanism of electron plasma instability in this structure. The instability in the plasmonic crystal formed in the transistor channel develops due to conversion of the kinetic energy carried by the drifting plasmons into electromagnetic energy. The conversion happens at the opposite sides of the gate fingers due to the asymmetry produced by the current flow and occurs through the gate finger fringing capacitances. The key feature of the proposed instability mechanism is the behavior of the plasma frequency peak and its width as functions of the dc current bias. At a certain critical value of the current, the plasma resonant peak with small instability increment experiencing redshift with increasing current changes to the blue shifting peak with large instability increment. This amplified mode switching (AMS) effect has been recently observed in graphene-interdigitated structures (S. Boubanga-Tombet *et al.*, Phys. Rev. X **10**, 031004 (2020)). The obtained theoretical results are in very good qualitative agreement with these experiments and can be used in future designs of the compact sources of THz EM radiation.

[1]gaizin@kbcc.cuny.edu

[†]shurm@rpi.edu

# I. INTRODUCTION

Application of sub-THz and THz technology ranging from 6G communications [1], homeland and hardware cyber security [2], industrial controls [3], VLSI testing [4] to biomedical applications [5], including cancer detection [6] and, possibly, even treatment [7] all require portable, efficient, and inexpensive detectors, and sources. The state-of-the-art THz sources include Schottky diode frequency multipliers [8], IMPATT diodes [9], Quantum Cascade lasers [10], Resonant Tunneling Diodes [11], and femtosecond laser systems for Time Domain Spectroscopy [12]. These sources cannot simultaneously satisfy the requirement of cost, efficiency, and portability. The plasmonic THz field effect transistors (TeraFETs) [13] have emerged as a prime candidate for compact, efficient, and inexpensive sub-THz and THz detectors and sources [14]. These devices use the excitation and instabilities of plasma waves in the transistor channels. Achieving efficient, portable, and tunable THz sources is an especially challenging and important technological problem. Different instability mechanisms including the Dyakonov-Shur (DS) instability [15] and transit time instability [16] have been explored theoretically and experimentally but the demonstrated plasmonic THz sources all need improvement in power and efficiency to enable the key application of the 6G communication [17]. To achieve these improvements, combining these transistors into arrays [18], using interdigitated structures [19, 20], and achieving a plasmonic boom instability [21, 22] have been proposed promising THz output powers in the 100 mW range.

In this paper, we explore a new mechanism of electron plasma instability in the current biased TeraFETs with dual interdigitated grating gates. This instability occurs in the plasmonic crystal formed in this structure. In the current biased electron channel, the drifting plasmons carry both electromagnetic (EM) and kinetic energy [23, 24]. We show that in the grating gated channel, the kinetic and EM energy flows exchange near the edges of the individual gate fingers due to fringing capacitances between the gate fingers and the channel. This may result in the partial conversion of the kinetic energy carried by drifting plasmon into the EM energy accompanied by the exponential growth of the plasma wave amplitude. The energy flow conversion leads to plasmonic instability in the plasmonic crystal cells. Coherence of the plasma wave over many elementary cells in the plasmonic crystal is maintained by the adjustment of the phases of the waves in the TeraFET sections so that the plasma instability develops in the entire channel. A similar effect takes place in a single gate TeraFET with periodically changing section width. In this case, conversion between kinetic and EM energies and the instability are achieved due to the "plasmonic stubs" playing the same role as the fringing capacitances in the interdigitated structures [24]. The instability is significantly improved in TeraFETs with asymmetric dual grating gates. In this geometry, the asymmetric boundary conditions in the elementary cells of the plasmonic crystal lead to a further increase of the instability increment for one preferred direction of the electron drift velocity. This effect is similar to the Dyakonov-Shur instability in asymmetric transistors [15]. The new and key feature of the proposed instability mechanism is the involvement of two plasmonic modes intersecting at a certain critical excitation dc current. At this current, a low increment plasmonic peak red shifting with the excitation dc current bias becomes a higher increment blue shifting peak. This amplified mode switching (AMS) effect has been observed in interdigitated graphene structures [25].

If the collisional and viscous damping of the plasma modes exceeded the plasma wave growth increment, our theory predicts the red and blue shifts of THz plasma frequency and impinging THz wave attenuation changes caused by a dc current flow. We show that these predictions are in good qualitative agreement with the recent measurements of the THz absorption in interdigitated graphene structures [25, 26]. We analyze the conditions necessary for the experimental observation of the instability and developing compact and efficient sources of THz radiation based on the mechanism analyzed in this paper.

The paper is organized as follows. In Section II, we derive the basic equations for the plasmonic crystal formed in the graphene TeraFET with an interdigitated grating gate. In Section III, we present solutions of the plasmon dispersion equations and analyze the plasma instability effect. Section IV compares the predictions of our theory with the available experimental data [25, 26]. Concluding remarks and a discussion of possible device applications are in Section V.

## II. BASIC EQUATIONS

We consider plasma oscillations in the 2D electron gas in graphene FET with the geometry shown in Fig. 1a. The graphene FET channel is sandwiched between two dielectric layers of thickness $d$. The back gate is used to uniformly change the electron density in the 2D channel, and the two top grating gates serve to provide the periodic modulation of the electron density with suitable profiles. These interdigitated dual-grating-gate graphene FETs were used recently to study various THz plasmonic effects in the graphene conducting channel with periodically modulated electron density [25-27]. A distinctive feature of this structure as compared to the FET with a single grating gate is the built-in asymmetry provided by the asymmetric positioning of the two top grating gates. As shown in a number of publications, this asymmetry is critical for the detection of THz EM radiation [27,28] and can potentially be used for a generation as well [25].

We restrict our consideration to the case when the equilibrium electron density profile in the graphene channel consists of periodically repeated segments 1 and 2 with high electron density $n_{01}$ and low electron density $n_{02}$ having lengths $L_1$ and $L_2$, respectively as shown in Fig. 1b. This density distribution can be easily achieved, for example by applying the proper gate biases to the back gate and one of the top grating gates while keeping another top grating gate at zero potential [25,27]. In this case, the electron system in the channel represents a periodic plasmonic medium forming a 1D plasmonic crystal [21, 22, 24, 29, 30]. We also assume that a dc current passes through the FET and is characterized by the electron drift velocities $v_{01}$ and $v_{02}$ in segments 1 and 2 with $n_{01}v_{01} = n_{02}v_{02}$. In the following calculations, we consider the graphene layer in the plane $z = 0$, choose the $x$-axis in the source-drain direction and assume that all characteristics of the 2D electron system depend on $x$-coordinate only.

Plasma waves in the 2D electron gas with periodically modulated electron density can be described in the hydrodynamic approximation provided that the electron-electron scattering length is the shortest characteristic length in the system, in particular, shorter than the lengths of the channel segments $L_{1,2}$ and the length of the transient region between segments, $\sim d$. In the hydrodynamic approximation, the system of equations for the electron density $n(x,t)$ and velocity $v(x,t)$ in the 2D electron fluid in graphene consists of the equation of continuity

$$\frac{\partial n}{\partial t} + \frac{\partial (nv)}{\partial x} = 0 \tag{1}$$

representing conservation of the number of electrons, and the Euler equation describing the electron fluid dynamics. The latter equation for the graphene electron fluid at arbitrary values of the fluid velocity $v(x,t) \leq v_F$, where $v_F$ is graphene Fermi velocity was derived in [31,32] and can be written in the following form

$$\frac{2-\beta^2}{2(1-\beta^2)}\frac{\partial v}{\partial t} + \frac{v}{2(1-\beta^2)}\frac{\partial v}{\partial x} + \frac{v}{2n}\frac{\partial n}{\partial t} + \frac{v_F^2}{2n}\frac{\partial n}{\partial x} + \frac{v_F(1-\beta^2)^{1/4}}{\sqrt{\pi\hbar}\sqrt{n}}eE^{ind} = 0, \tag{2}$$

where $\beta(x,t) = \frac{v(x,t)}{v_F}$ is the dimensionless local drift velocity, $E^{ind}$ is the self-consistent electric field induced by the plasma charge fluctuations, $-e$ is the electron charge. In Eq. (2), we neglected the collision term due to random scattering of plasma fluctuations assuming that $\omega\tau \gg 1$, where $\omega$ is the plasmon frequency and $\tau$ is the characteristic scattering time. The Euler equation (2) is significantly simplified in the limit of the small local drift velocities when $\beta^2 \ll 1$. In this limit, we linearize Eqs. (1) and (2) for the small fluctuations of electron density and the local drift velocity $n(x,t) = n_0 + \delta n_{q\omega}e^{-iqx+i\omega t}$ and $v(x,t) = v_0 + \delta v_{q\omega}e^{-iqx+i\omega t}$. In the quasi-static limit, the induced electric field $\delta E_{q\omega}^{ind}$ and the charge fluctuation $-e\delta n_{q\omega}$ are connected via the local gate capacitance $C = \frac{\epsilon\epsilon_0}{d}$ as

$$\delta E_{q\omega}^{ind} = -\frac{iqe\delta n_{q\omega}}{C}, \tag{3}$$

where $\epsilon$ is the dielectric constant of the gate dielectric. With these approximations, Eqs. (1) and (2) yield the following system of linear equations for the Fourier harmonics $\delta n_{q\omega}$ and $\delta v_{q\omega}$

$$\begin{cases} qn_0\delta v_{q\omega} - (\omega - qv_0)\delta n_{q\omega} = 0 \\ \omega\delta v_{q\omega} - q\left(\frac{e^2 v_F}{\sqrt{\pi\hbar}\sqrt{n_0}C} + \frac{v_F^2}{2n_0}\right)\delta n_{q\omega} = 0 \end{cases} \tag{4}$$

This system has a non-trivial solution if and only if $\omega = v_p^{(\pm)}q$, where

$$v_p^{(\pm)} = \frac{v_0}{2} \pm v_p, \quad v_p = \sqrt{\frac{v_F^2}{2} + \frac{e^2\sqrt{n_0}dv_F}{\sqrt{\pi}\epsilon\epsilon_0\hbar}} \tag{5}$$

Here, $v_p^{(\pm)}$ are velocities of the plasmons traveling downstream (+) and upstream (-) in the graphene electron fluid drifting with velocity $v_0$, and $v_p$ is the plasma wave velocity in the absence of drift [33]. The general solution of Eq. (4) can be expressed in terms of the total plasmonic current in the channel of width $W$: $I_{q\omega} = -e(n_0\delta v_{q\omega} + v_0\delta n_{q\omega})W$ and the electric potential in the channel $V_{q\omega} = -\frac{e\delta n_{q\omega}}{C}$. For the plasmon of frequency $\omega$ we obtain

$$\begin{cases} I_\omega(x) = I_1 e^{-iq_1 x} + I_2 e^{-iq_2 x} \\ V_\omega(x) = \frac{1}{CW}\left( \frac{I_1}{v_p(1+\beta_0/2)} e^{-iq_1 x} - \frac{I_2}{v_p(1-\beta_0/2)} e^{-iq_2 x} \right) \end{cases}, \quad (6)$$

where $\beta_0 = \frac{v_0}{v_p}$, $q_{1,2} = \frac{\omega}{v_0/2 \pm v_p}$, and constants $I_{1,2}$ are determined by the boundary conditions.

Equation (6) describes the spatial distribution of the plasmonic current $I_\omega$ and potential $V_\omega$ in the gated graphene channel at the frequency $\omega$. It follows from Eq. (6) that the values of $I_\omega$ and $V_\omega$ at the opposite ends of the gated plasmonic cavity of length $\ell$ ($x = 0, \ell$) are connected by the transfer matrix $\hat{t}$:

$$\begin{pmatrix} V_\omega(0) \\ I_\omega(0) \end{pmatrix} = \hat{t} \begin{pmatrix} V_\omega(\ell) \\ I_\omega(\ell) \end{pmatrix} \quad (7)$$

$$\hat{t} = \begin{pmatrix} \cos\frac{\omega \ell}{v_p} - i\beta_0 \sin\frac{\omega \ell}{v_p} & iZ_0 \sin\frac{\omega \ell}{v_p} \\ \frac{i}{Z_0} \sin\frac{\omega \ell}{v_p} & \cos\frac{\omega \ell}{v_p} + i\beta_0 \sin\frac{\omega \ell}{v_p} \end{pmatrix} \quad (8)$$

where $Z_0 = \frac{1}{WCv_p}$ is the characteristic impedance of the plasmonic transmission line representing the cavity [29].

Equations (6)-(8) describe the plasmon dynamics in the gated segments 1 and 2 of the graphene channel in Fig. 1. To obtain dispersion relation for the 1D plasmonic crystal formed in the graphene FET these solutions should be matched at the boundaries between the segments. The customary boundary conditions include continuity of the plasmonic current in the transistor channel. However, in the transistor with the grating gate, this current continuity breaks down at the boundary between the gated and ungated (back-gated) parts of the channel because part of the ac plasmonic current branches off into the gate due to fringing capacitive coupling between the gate fingers and the ungated part of the electron channel. If the lengths of the gate finger $L_1$ and the ungated (back-gated) section of the channel $L_2$ are much larger than the distance between them ($\sim d$) this fringing capacitance $C_b$ is essentially the capacitance of the co-planar capacitor: $C_b = \alpha \epsilon \epsilon_0 W$ where $\alpha \sim 1$ is the geometric factor depending on $L_{1,2}$ and $d$ as shown in the Appendix. The presence of this capacitance largely went unnoticed in the previous studies of the plasmonic transistor structures with the grating gate. The second grounded grating gate with a length much shorter than $L_1$ has no direct effect on the electron gas in the channel but changes the fringing capacitance on one side of the gated plasmonic cavity in comparison with the other one making the cavity boundaries asymmetric. As we show in the next section this asymmetry strongly affects the plasma instability in the system.

The electric circuit diagram for the dual-grating-gate graphene channel is shown in Fig. 1c. In this Figure, plasmonic transmission lines TL1 and TL2 describe the top-gated and back-gated cavities 1 and 2, respectfully, and the impedances $C_{b1}$ and $C_{b2}$ correspond to the fringing capacitances at the boundaries between the cavities, see Appendix. The presence of the second grounded gate changes the fringing capacitance on one side so that $C_{b1} \neq C_{b2}$, and we assume that $C_{b2} = \gamma C_{b1}$ where the asymmetry factor $\gamma$ is the fitting parameter of the model. The transfer matrix connecting the voltage $V_\omega$ and the plasmonic current $I_\omega$ across the capacitors $C_{b1,2}$ can be found from the continuity of the voltage and conservation of the total current as

$$\hat{s}(Z_{b1,2}) = \begin{pmatrix} 1 & 0 \\ \frac{1}{Z_{b1,2}} & 1 \end{pmatrix} \qquad (9)$$

where $Z_{b1,2} = \frac{1}{i\omega C_{b1,2}}$.

Using Eqs. (7)-(9) one can connect the values of $V_\omega$ and $I_\omega$ at the opposite ends of the plasmonic crystal elementary cell at $x = 0$ and $x = L = L_1 + L_2$ :

$$\begin{pmatrix} V_\omega(0) \\ I_\omega(0) \end{pmatrix} = \hat{t}_1 \hat{s}(Z_{b1}) \hat{t}_2 \hat{s}^{-1}(Z_{b2}) \begin{pmatrix} V_\omega(L) \\ I_\omega(L) \end{pmatrix}, \qquad (10)$$

where matrices $\hat{t}_1$ and $\hat{t}_2$ defined in Eq. (8) correspond to the gated and back-gated cavities with different electron densities. In the general form, the dispersion relation for the 1D drifting plasmonic crystal can be found from Eq. (10) and the Bloch boundary conditions $(V_\omega(L), I_\omega(L)) = e^{-ikL}(V_\omega(0), I_\omega(0))$ as [24]

$$\det \hat{T} - e^{ikL} Tr \hat{T} + e^{2ikL} = 0, \qquad (11)$$

where $\hat{T} = \hat{t}_1 \hat{s}(Z_{b1}) \hat{t}_2 \hat{s}^{-1}(Z_{b2})$ and $k \in \left[-\frac{\pi}{L}, \frac{\pi}{L}\right]$ is the plasmon Bloch wave vector. Substituting expressions for the *t*- and *s*-matrices given by Eqs. (8) and (9) into Eq. (11) after some lengthy but straightforward calculations we finally obtain

$$\cos\left(kl + \frac{1}{2}M_1\theta_1 + \frac{1}{2}M_2\theta_2\right) = \cos\theta_1 \cos\theta_2 - \frac{1}{2}\left[\eta^{1/4} + \eta^{-1/4} + \frac{1}{2}\left(\frac{1}{Z_{b1}} + \frac{1}{Z_{b2}}\right)(M_1 Z_{02} - M_2 Z_{01}) - \frac{Z_{01}Z_{02}}{Z_{b1}Z_{b2}}\right] \sin\theta_1 \sin\theta_2 + \frac{i}{2}\left(\frac{1}{Z_{b1}} - \frac{1}{Z_{b2}}\right)(Z_{02} \cos\theta_1 \sin\theta_2 + Z_{01} \cos\theta_2 \sin\theta_1) \qquad (12)$$

Here, $\eta = \frac{n_{02}}{n_{01}}$, $\theta_i = \frac{\omega L_i}{v_{pi}}$, $M_i = \frac{v_{0i}}{v_{pi}}$, $Z_{0i} = \frac{1}{WCv_{pi}}$, where the index $i = 1,2$ refers to the top-gated and back-gated cavities, respectively. Equation (12) determines the energy spectrum of the drifting 1D plasmonic crystal in the graphene transistor with the grating gate. This equation is analyzed in the next section.

### III. RESULTS

In this section, we use the dispersion Eq. (12) to analyze the energy spectrum of the drifting plasmonic channel.

In Fig. 2, we plot the plasmonic band spectrum $\omega(k) = \omega'(k) + i\omega''(k)$, $-\pi \leq kL \leq \pi$, found from Eq. (12) at different values of the drift velocity. In this calculation, we assumed $L_1 = L_2 = L/2$, $\frac{d}{L} = 0.005$, $\eta = 0.1$, and the asymmetry factor $\gamma = 2$. The spectra shown on different panels of Fig. 2 correspond to the different drift velocities described by the Mach number $M$ in the back-gated cavities, $M = \frac{v_{02}}{v_{p2}}$. The plots of $\omega'(k)$ demonstrate that the discrete plasmon energy levels in the cavities are broadened into the energy bands as expected. At finite $M$ the bands are shifted because of the Doppler effect.

The most remarkable result is the appearance of the imaginary part of the plasma frequency at $M \neq 0$. These finite values of $\omega''$ added to the positive imaginary part of the plasma frequency due to ordinary collisional damping may either increase (at $\omega'' > 0$) or decrease (at $\omega'' < 0$) the overall damping of the drifting band plasmons [34]. If $\omega'' < 0$ and $|\omega''|$ exceeds the collisional damping the plasma instability is developed. The physical reason for finite $\omega''$ in the energy spectrum of the plasma excitations drifting in the plasmonic medium comprised of periodically repeated different sections was discussed in Refs. [23,24]. The total power flow in the drifting plasmon consists of EM and kinetic powers. If the total power flow at both ends of the crystal elementary cell in the channel is the same the plasmon remains stable, and $\omega''=0$ ($\omega''<0$ if the collisional damping is accounted for). The presence of the fringing capacitors breaks this continuity because part of the plasmonic current branches off the channel into the gate and vice versa. The most striking example of this behavior is the Dyakonov-Shur instability in the transistor channel with asymmetric boundaries [15]. The sign of $\omega''$ depends on the phase-matching conditions of the plasma waves in different elementary cells, i.e., $\omega''$ can have opposite signs in different finite intervals of the Bloch wave vector values for one plasma mode. It can also change the sign for different plasma modes at the same value of the Bloch wave vector [24,30]. This qualitative explanation is confirmed by the results presented in Fig. 2. Our calculations (not shown here) also demonstrate that $\omega''$ disappears in the absence of the fringing capacitances ($C_{b1,2} = 0$) at any value of $M$.

Plasma modes at the center of the Brillouin zone ($k = 0$) present special interest because these modes are probed by the external EM radiation incident on the transistor with the grating gate. In Fig. 3, we plot the evolution of these modes with increasing electron density $n_{01}$ proportional to the grating gate voltage. In this calculation, we take $M = 0$, so that $\omega'' = 0$, and $\gamma = 2$. Without density modulation ($n_{01} = n_{02}$) the plasma modes at the center of the Brillouin zone have frequencies $\omega_n = \frac{2\pi v_p}{L} n, n = 1,2, ...$, where $v_{p1} = v_{p2} = v_p$ is defined in Eq. (5) (the empty lattice limit). When $n_{01}$ increases the two types of quantized plasma modes localized in either the top-gated or the back-gated cavities emerge. The frequencies of the back-gated plasma modes do not depend on $n_{01}$ and presented by nearly horizontal lines in Fig. 3. The frequencies of the top-gated modes increase with $n_{01}$ as $\omega \sim n_{01}^{1/4}$. This dependence is characteristic of the gated plasma modes in graphene [35]. Due to mode coupling the final plasmon spectrum is formed after multiple anti-crossings of the original modes with splitting depending on the mode parity. The final plasma modes are spread over multiple elementary cells and cannot be assigned to only the top-gated or the back-gated interacting plasmonic cavities.

In Fig. 4, we plot the frequencies $\frac{\omega L}{v_{p2}}$ of several low-lying plasma modes at the center of the Brillouin zone as a function of the dimensionless drift velocity $\frac{v_{02}}{v_F}$ at different values of the asymmetry factor $\gamma = \frac{C_{b2}}{C_{b1}}$. In this calculation we used $v_F = 1 \times 10^6 m/s$ and $n_{01} = 1 \times 10^{16} m^{-2}$. All other parameters are the same as in Fig. 2. The most noticeable feature of this dependence is either a red or blue shift of the mode frequencies with changing drift velocity. Physically, the shift occurs due to shifting the plasmon dispersion curves in the $k$-space with increasing $M$, cf. Fig. 2. It has the same nature as the red or blue shift of the frequencies of the upstream and downstream plasmons in Eq. (5) due to the Doppler effect.

These shifts cause additional anti-crossings of different modes shown in Fig. 4. Near the anti-crossing points there are two plasma modes with the same frequency $\omega$ at two different values of the drift velocity. One of these modes is subject to the redshift and another one to the blueshift with increasing drift velocity. This type of plasma mode behavior was recently observed in the experiment [25] and will be discussed in more detail in the next section.

The imaginary part of the plasma frequency $\omega''$ peaks near the anti-crossing points making these points most susceptible to the non-monotonic changes in the plasmon damping or instability. The absolute value of $\omega''$ depends on the asymmetry of the fringing capacitances at the edges of the top-gated cavities characterized by the parameter $\gamma$ and also on the direction of the drift velocity $v_0$ in asymmetric structures. Comparison of the results obtained for the symmetric structures ($\gamma = 1$) in Fig. 4a with that for the asymmetric structures ($\gamma = 10$) in Figs. 4b and 4c shows that in the asymmetric structures the absolute value of $\omega''$ increases though the plasmon frequency $\omega'$ remains essentially the same. In asymmetric structures, the value of $\omega''$ also depends on the direction of the drift velocity. It follows from the comparison on Figs. 4b and 4c where on both Figures $\gamma = 10$ but $v_0 > 0$ in Fig. 4b and $v_0 < 0$ in Fig. 4c.

The analysis in this section demonstrates a number of new promising features in the plasmon spectrum in the THz graphene transistor with the interdigitated grating gate which can make this system attractive for experimental studies for both fundamental and applied purposes.

## IV. DISCUSSION

The interaction of the THz EM radiation with plasmons in the graphene transistors with interdigitated grating gates was recently studied experimentally in several papers [25-27]. In this section, we analyze the experimentally verifiable predictions of our theoretical model and compare the results with current experimental data.

In Fig. 5, we plot the frequencies $f$ and the instability increments/decrements $\omega''$ of the plasma modes at the center of the Brillouin zone as a function of the electron drift velocity $v_{02}/v_F$ in the back-gated sections of the graphene channel. These modes are excited by an external THz EM wave incident on the periodic transistor structure. In this calculation, we used the grating gate period $L = 4\mu m$, the thickness of the gate dielectric $d = 20nm$, and the dielectric constant $\epsilon = 4.5$. We also assumed that the electron densities in the top-gated and back-gated sections of the channel are $n_{01} = 1 \times 10^{16} \, m^{-2}$ and $n_{02} = 0.1 \times 10^{16} \, m^{-2}$, respectively, the Fermi velocity in graphene $v_F = 1 \times 10^6 \, m/s$ and the asymmetry factor $\gamma = 2$. The strength of the resonant interaction between the plasma mode of frequency $\omega$ and an external EM wave is determined by the quality factor $Q = \omega\tau$ where $\tau$ is the mode relaxation time. For a typical experimental relaxation time $\tau \sim 0.1 ps$, the first six modes (dashed gray lines in Fig. 5) have the quality factor $Q < 1$, and these modes are suppressed due to excessive damping. For illustrative purposes, we chose $\tau = 0.09 ps$ so that the quality factor $Q = 1$ corresponds to the threshold plasma frequency $f_s = 1.77\ THz$ (black horizontal line in Fig. 5). In this configuration, the frequency of the lowest active plasma mode (the red line in Fig. 5) first experiences the redshift when the electron drift velocity becomes finite and increases. The $Q$-factor of the mode decreases and at some value of the drift velocity $v_{02}^A$ reaches the critical value $Q = 1$, see Fig. 5. At $v_{02} > v_{02}^A$ the height of the resonant peak effectively reduces to zero due to increased damping. However, when the drift velocity continues to increase the resonant peak re-emerges at a larger value of the drift velocity $v_{02}^B$, see Fig. 5. The frequency of this re-emerged mode experiences the blueshift with increasing drift velocity, and its $Q$-factor increases as well. This interesting switching behavior together with the finite gap with no plasmonic absorption peak at $v_{02}^A < v_{02} < v_{02}^B$ was recently observed in Ref. [25].

In Ref. [25], the authors also reported decreased resonant absorption of external EM radiation in the plasmonic crystal in comparison with the same absorption that occurred without the electron density modulation by the grating gate. This effect was interpreted as an amplification of the THz EM radiation by the plasma excitations in the density-modulated electron system. This unusual behavior can be explained in our theoretical model as follows.

In Fig. 5, we also plotted the imaginary part of the plasma frequency $\omega''$ as a function of the drift velocity. The value of $\omega''$ for the active mode (the blue line in Fig. 5) is negative at almost all values of the drift velocity indicating that this mode should be subject to instability with the instability increment $|\omega''|$ depending on the drift velocity. The instability is countered by the plasmon damping $\frac{1}{\tau}$. Since $\frac{1}{\tau} \gg |\omega''|$ the plasma mode experiences damping but with the effective relaxation time $\tau_{eff}$ defined as $\frac{1}{\tau_{eff}} = \frac{1}{\tau} - \frac{1}{|\omega''|}$. The absorption of an external EM wave by the plasmons is determined by this effective damping and therefore depends on the drift velocity. It follows from Eq. (12) that without density modulation ($n_{01} = n_{02}$) $\omega'' = 0$ at any drift velocities. In the density-modulated system ($n_{01} > n_{02}$) at finite drift velocities $\omega'' < 0$, and the overall damping rate $\frac{1}{\tau_{eff}}$ and absorption are reduced. This AMS effect was observed in the experiment [25].

As shown in Fig. 5, $\omega''$ changes non-monotonically with the drift velocity reaching the maximum negative value somewhere between $v_{02}^A$ and $v_{02}^B$ where the plasma frequency switches its behavior from red shifting to blue shifting. However, the peak in $\omega''$ is not symmetric with the values of $|\omega''|$ noticeably larger at large drift velocities where the plasma frequency experiences the blueshift. This can explain why the negative changes in absorption were observed only at the drift velocities corresponding to the blue shift of the frequency. This qualitative explanation does not account for the changes in absorption due to increased electron density in the density-modulated channel but the latter changes do not depend on the drift velocity and are not responsible for the different behavior of the absorption in the regions with opposite frequency shifts. No absorption decrease with the dc current excitation was reported in the graphene TeraFETs with symmetric dual grating gates [26]. This result can be explained by the very small values of the instability increments in symmetric structures with asymmetry factor $\gamma = 1$ as shown in Fig. 4a.

The above analysis validates the developed theoretical model. The presented theory qualitatively explains the experimental results [25, 26]. It also provides general quantitative agreement with the experiment in terms of the predicted plasma frequencies and the used range of the drift velocities, all with the Mach number $M < 1$. The numerical estimates were done for a system with the material parameters and geometry close to the values used in the experiment. The gate biasing diagram shown in Fig. 1a differs from the biasing scheme used in the experiment [25] where both top grating gates were utilized to control the electron density in the channel. This may affect the values of the fringing capacitances and change numerical estimates. However, since the equilibrium electron density profile remains the same regardless of the biasing scheme, the theoretical results will not change qualitatively. Also, in the experiment [25] the length of the top-gated and back-gated cavities, $L_1$ and $L_2$ in Fig. 1a, were different. This leads to additional dependence of the plasma resonant effects on the ratio $L_1/L_2$ which is not captured in our model where $L_1 = L_2$ was assumed. The theoretical estimates of the drift velocity necessary to observe the AMS behavior of the plasma frequencies are larger than the values estimated in the experiment [25]. This discrepancy could be due to uncertainties in the values of some material parameters we used in our calculations, in particular, the Fermi velocity in graphene, the dielectric constant, and the thickness of the gate dielectric in the experimental samples.

## V. CONCLUDING REMARKS

The presented theory is qualitatively confirmed by the available experimental data [25] and can be used for designing compact and efficient THz sources based on the TeraFETs with interdigitated grating gates. Plasmonic TeraFET efficiently generates the EM radiation when the instability increment of the plasma modes exceeds the plasmon damping rate. So far, this condition has not been reached in experimental studies. Our theory suggests several paths to overcome this limitation. The calculated plasma mode increment increases together with the plasma frequency and can exceed damping in samples with sufficiently large electron density and/or smaller grating period. The increment also significantly increases in the samples with asymmetric fringing capacitances or stubs [24]. The asymmetry could be further enhanced by the gate finger design such as threshold voltage profiling under the gates [36] increasing the asymmetry factor $\gamma$. The power of the THz EM signal generated in transistors with the grating gate is much larger than in a single-gate transistor because of the coherent addition of the signals generated in each elementary cell of the plasmonic crystal formed in the transistor channel. In this paper, we considered the interdigitated TeraFET graphene structure as an example of the suggested instability effect because of the available experimental data. However, this approach could equally apply to conventional TeraFET materials, such as Si [37], InGaAs [38], or even p-diamond [39], and to other structures, with interdigitated rings being most promising for detection because of the contactless THz excitation inducing fairly high magnetic fields [40]. Although the developed theory applies to both semiconductor and graphene structures, the record high mobilities in graphene at room temperature make it a preferable material for observation of the instability effects.

In summary, we developed a theory of collective plasma oscillations in the graphene transistor with interdigitated dual-grating gate. We demonstrated that in the current biased transistor, some plasma modes become unstable and decrease the attenuation of the EM wave interacting with the channel. The plasma frequencies of different plasma modes can experience either red or blue shift with increasing electron drift velocity. A single plasma mode can switch its behavior from red-shifted to blue-shifted with increasing drift velocity. These two regimes are separated by the interval of the drift velocities where the plasma mode is completely suppressed. The theoretical results obtained in this paper are in very good qualitative agreement with recent experiments [25,26] and have wide-ranging applications for future designs of the new compact sources of the THz EM radiation.

## VI. ACKNOWLEDGMENTS

The work at RPI was supported by AFOSR (contract number FA9550-19-1-0355)

## APPENDIX

In this Appendix, for reference purposes, we estimate the fringing capacitance between the grating gate finger of length $L_1$ and the adjacent ungated part of the 2D electron channel of length $L_2$. The schematic diagram in Fig. 6 (a) shows the geometry modeling the boundary between the gated and ungated parts of the electron channel. It includes the 2D electron layer positioned at $y = 0$ and $-\infty \leq x \leq \infty$ and the metal gate positioned at $y = d$ and $-\infty \leq x \leq 0$. The gate finger and the 2D layer are held at electric potentials $V = V_0$ and $V = 0$, respectively. The relevant electrostatic problem is solved using the conformal mapping technique for the solution of two-dimensional potential problems [41]. Complex electric potential $W(x,y) = u + iV(x,y)/V_0$ in the whole $x$-$y$ plane can be found by the Schwartz-Christoffel mapping method as [41]

$$z = \frac{d}{\pi}(1 + \pi W + e^{\pi W}),  \qquad (A1)$$

where $z = x + iy$, and $u$ is a real parameter ($-\infty \leq u \leq \infty$) defining parametric equations of the equipotential lines ($V = const$) in Eq. (A1). Some of these lines are shown in Fig. 6 (a). At $V = 0$, it follows from Eq. (A1) that $y = 0$ and

$$x = \frac{d}{\pi}(1 + \pi u + e^{\pi u}) \qquad (A2)$$

The charge density $\sigma(x)$ in the 2D layer can be found from Eq. (A1) as

$$\sigma(x) = \epsilon\epsilon_0 V_0 \left|\frac{dW}{dz}\right|_{y=0} = \frac{\epsilon\epsilon_0 V_0}{d}\frac{1}{1+e^{\pi u}}, \qquad (A3)$$

where $\epsilon$ is the dielectric constant of the surrounding medium. Equations (A2) and (A3) allow evaluation of the charge $Q$ accumulated on the capacitor plates by integrating Eq. (A3) along $x$-axis. Since the charge density in Eq. (A3) exponentially decreases at $x \to \infty$ and approaches constant value at $x \to -\infty$ over characteristic distances $\sim d$ the integration can be limited to the interval $-L_1 \leq x \leq L_2$ provided that $L_{1,2} \gg d$. In this limit, as it follows from Eq. (A2) points $x_1 = -L_1$ and $x_2 = L_2$ on the $x$-axis correspond to the points $u_1 = -\frac{L_1}{d}$ and $u_2 = \frac{1}{\pi}\ln\frac{\pi L_2}{d}$ on the $u$-axis. Then

$$Q = W\int_{-L_1}^{L_2}\sigma(x)dx = W\int_{u_1}^{u_2}\sigma(u)\frac{dx}{du}du = C_{tot}V_0, \qquad (A4)$$

where $W$ is the channel width and

$$C_{tot} = \frac{\epsilon\epsilon_0}{d}L_1 W + \frac{\epsilon\epsilon_0 W}{\pi}\ln\frac{\pi L_2}{d} \qquad (A5)$$

is the total capacitance between the gate and the 2D electron layer. The first term in Eq. (A5) corresponds to the capacitance $C_{pl} = \frac{\epsilon\epsilon_0}{d}L_1 W$ of the plate capacitor formed between the gate and the electron layer underneath the gate and the second term represents the fringing capacitance

$$C_b = \alpha\epsilon\epsilon_0 W, \quad \alpha = \frac{1}{\pi}\ln\frac{\pi L_2}{d} \qquad (A6)$$

The equivalent electric circuit corresponding to Fig. 6(a) is shown in Fig. 6 (b). These results are used in our calculations in Sections II and III.

# REFERENCES

[1] M. Shur, G. Aizin, T. Otsuji, V. Ryzhii, *Plasmonic Field-Effect Transistors (TeraFETs) for 6G Communications*, Sensors **21**, 7907 (2021).

[2] N. Akter, N. Pala, W. Knap, M. Shur, *THz Plasma Field Effect Transistor Detectors*, in Fundamentals of Terahertz Devices and Applications, edited by D. Pavlidis (Wiley, 2021), Chap. 8

[3] M. Naftaly, N. Vieweg, A. Deninger, *Industrial Applications of Terahertz Sensing: State of Play*, Sensors **19**, 4203 (2019).

[4] N. Akter, M. R. Siddiquee, M. Shur and N. Pala, *AI-Powered Terahertz VLSI Testing Technology for Ensuring Hardware Security and Reliability*, IEEE Access **9**, 64499 (2021).

[5] M. Shur and X. Liu, *Biomedical applications of terahertz technology*, in Proc. SPIE **11975**, Advances in Terahertz Biomedical Imaging and Spectroscopy, 1197502 (2022).

[6] P. Doradla, C. Joseph, and R. H. Giles, *Terahertz endoscopic imaging for colorectal cancer detection: Current status and future perspectives*, World J. Gastrointest. Endosc. **9**, 346 (2017).

[7] L. V. Titova, A. K. Ayesheshim, A. Golubov, R. Rodriguez-Juarez, R. Woycicki, F. A. Hegmann, and O. Kovalchuk, *Intense THz pulses down-regulate genes associated with skin cancer and psoriasis: a new therapeutic avenue?*, Sci. Rep. **3**, 2363 (2013).

[8] E. Schlecht, J. Siles, C. Lee, R. Lin, B. Thomas, G. Chattopadhyay, I. Mehdi, *Schottky Diode Based 1.2 THz Receivers Operating at Room-Temperature and Below for Planetary Atmospheric Sounding*, IEEE Trans. Terahertz Sci. and Technol. **4**, 661 (2014).

[9] A. Acharyya, J. P. Banerjee, *Prospects of IMPATT devices based on wide bandgap semiconductors as potential terahertz sources*, Appl Nanosci. **4**, 1 (2014).

[10] B. Wen, and D. Ban, *High-temperature terahertz quantum cascade lasers*, Prog. Quantum Electron. **80**, 100363 (2021).

[11] H. Sugiyama, S. Suzuki, and M. Asada, *Room-temperature Resonant-tunneling-diode Terahertz Oscillator Based on Precisely Controlled Semiconductor Epitaxial Growth Technology*, NTT Tech. Rev. **9**, No. 10 (2011).

[12] J. Neu and C. A. Schmuttenmaer, *Tutorial: An introduction to terahertz time-domain spectroscopy (THz-TDS)*, J. Appl. Phys. **124**, 231101 (2018).

[13] Y. Salamin, I-C. Benea-Chelmus, Y. Fedoryshyn, W. Heni, D. L. Elder, L. R. Dalton, J. Faist, and J. Leuthold, *Compact and ultra-efficient broadband plasmonic terahertz field detector*, Nat. Commun. **10**, 5550 (2019).

[14] M. S. Shur, *Terahertz Plasmonic Technology*, IEEE Sens. J. **21**, 12752 (2021).

[15] M. Dyakonov and M. Shur, S*hallow Water Analogy for a Ballistic Field Effect Transistor: New Mechanism of Plasma Wave Generation by DC Current,* Phys. Rev. Lett. **71**, 2465 (1993).

[16] V. Ryzhii, A. Satou, and M. Shur, *Transit Time Mechanism of Plasma Instability in High Electron Mobility Transistors,* Phys. Status Solidi A **202**, R113 (2005).

[17] M. Shur, S. Rudin, G. Rupper, F. Dagefu and M. Brodsky *TeraFETs for Terahertz Communications*, IEEE Photon. Soc. Newsletter **33**, 4 (2019).

[18] D. M. Yermolayev, E. A. Polushkin and S. Yu. Shapoval, V. V. Popov, K. V. Marem'yanin, V. I. Gavrilenko, N. A. Maleev and V. M.V. E. Zemlyakov, V. I. Yegorkin and V. A. Bespalov, A. V. Muravjov, S. L. Rumyantsev, and M. S. Shur, *Detection of Terahertz Radiation by Dense Arrays of InGaAs Transistors*, Int. J. High Speed Electron. Syst. **24**, 1 (2015).

[19] T. Otsuji, Y. M. Meziani, T. Nishimura, T. Suemitsu, W. Knap, E. Sano, T. Asano, and V. V. Popov, *Emission of Terahertz Radiation from Dual Grating Gate Plasmon-resonant Emitters Fabricated with InGaP/InGaAs/GaAs Material Systems,* J. Phys. Condens. Matter **20**, 384206 (2008).

[20] K. Tamura, D. Ogiura, K. Suwa, H. Fukidome, A. Satou, Y. Takida, H. Minamide, and T. Otsuji *Terahertz Detection by an Asymmetric Dual-Grating-Gate Graphene FET*, in *Proceedings of the 46th International Conference on Infrared, Millimeter and Terahertz Waves (IRMMW-THz), Chengdu, China* (IEEE, New York, 2021).

[21] V. Y. Kachorovskii and M. S. Shur, *Current-Induced Terahertz Oscillations in Plasmonic Crystal,* Appl. Phys. Lett. **100**, 232108 (2012).

[22] G. R. Aizin, J. Mikalopas, and M. Shur, *Current-Driven Plasmonic Boom Instability in Three-Dimensional Gated Periodic Ballistic Nanostructures,* Phys. Rev. B **93**, 195315 (2016).

[23] A. S. Petrov and D. Svintsov, *Perturbation theory for two-dimensional hydrodynamic plasmons,* Phys. Rev. B **99**, 195437 (2019).

[24] G. R. Aizin, J. Mikalopas, and M. Shur, *Plasmonic instabilities in two-dimensional electron channels of variable width,* Phys. Rev. B, **101**, 245404 (2020).

[25] S. Boubanga-Tombet, W. Knap, D. Yadav, A. Satou, D. B. But, V. V. Popov, I. V. Gorbenko, V. Kachorovskii, and T. Otsuji, *Room-Temperature Amplification of Terahertz Radiation by Grating-Gate Graphene Structures,* Phys. Rev. X **10**, 031004 (2020) and references therein.

[26] S. Boubanga-Tombet, A. Satou, D. Yadav, D. B. But, W. Knap, V. V. Popov, I. V. Gorbenko, V. Kachorovskii, and T. Otsuji, *Paving the Way for Tunable Graphene Plasmonic THz Amplifiers,* Front. Phys. **9**, 726806 (2021).

[27] P. Olbrich, J. Kamann, M. Konig, J. Munzert, L. Tutsch, J. Eroms, D. Weiss, M.-H. Liu, L. E. Golub, E. L. Ivchenko, V. V. Popov, D. V. Fateev, K. V. Mashinsky, F. Fromm, T. Seyller, and S. D. Ganichev, *Terahertz Rachet Effects in Graphene with a Lateral Superlattice,* Phys. Rev. B, **93**, 075422 (2016).

[28] P. Faltermeier, P. Olbrich, W. Probst, L. Schell, T. Watanabe, S. A. Boubanga-Tombet, T. Otsuji, and S. D. Ganichev, *Helicity Sensitive Terahertz Radiation Detection by Dual-Grating-Gate High Electron Mobility transistors,* J. Appl. Phys. **118**, 084301 (2015).

[29] G. R. Aizin and G. C. Dyer, *Transmission line theory of collective plasma excitations in periodic two-dimensional electron systems: Finite plasmonic crystals and Tamm states,* Phys. Rev. B **86**, 235316 (2012).

[30] A. S. Petrov, D. Svintsov, V. Ryzhii, M. S. Shur, *Amplified-reflection plasmon instabilities in grating-gate plasmonic crystals,* Phys. Rev. B **95**, 045405 (2017).

[31] U. Briskot, M. Schutt, I. V. Gornyi, M. Titov, B. N. Narozhny, and A. D. Mirlin, *Collision-dominated Nonlinear Hydrodynamics in graphene,* Phys. Rev B **92**, 115426 (2015).

[32] J. Crabb, X. Cantos-Roman, J. M. Jornet, and G. R. Aizin, *Hydrodynamic theory of the Dyakonov-Shur instability in graphene transistors,* Phys. Rev. B **104**, 155440 (2021).

[33] D. Svintsov, V. Vyurkov, V. Ryzhii, and T. Otsuji, *Hydrodynamic electron transport, and nonlinear waves in graphene,* Phys. Rev. B **88**, 245444 (2013).

[34] D. Veksler, F. Teppe, A. P. Dmitriev, V. Y Kachorovskii, W. Knap, and M. S. Shur, *Detection of Terahertz Radiation in gated Two-Dimensional Structures Governed by DC Current,* Phys. Rev. B **73**, 125328 (2006).

[35] A. N. Grigorenko, M. Polini, and K. S. Novoselov, *Graphene Plasmonics,* Nature Photonics **6**, 749 (2012).

[36] Y. Zhang and M. Shur, *TeraFET terahertz detectors with spatially non-uniform gate capacitances*, Appl. Phys. Lett. **119**, 161104 (2021).

[37] X. Liu, T. Ytterdal, and M. Shur, *Plasmonic FET Terahertz Spectrometer*, IEEE Access **8**, 56039 (2020).

[38] M. Shur, J. Mikalopas and G. R. Aizin, *Compact Design Models of Cryo and Room Temperature Si MOS, GaN, InGaAs, and p-diamond HEMT TeraFETs*, in *Proceedings of the 2020 IEEE Radio and Wireless Symposium (RWS), San Antonio, TX, USA* (IEEE, New York, 2020), p. 209.

[39] Y. Zhang and M. S. Shur, *p-Diamond, Si, GaN, and InGaAs TeraFETs*, IEEE Trans. Electron Devices **67**, 4858-4865 (2020).

[40] G. Aizin, J. Mikalopas, and M. S. Shur, *Giant Inverse Faraday Effect in Plasmonic Crystal Ring*, Opt. Express **30**, 13733 (2022).

[41] P. Morse and H. Feshbach, *Methods of Theoretical Physics*, Part II (McGraw-Hill, New York, 1953), Chap. 10.

# FIGURE CAPTIONS

FIG. 1. (a) Schematics of the graphene FET with a dual asymmetric grating gate; (b) Spatial profile of the 2D electron density in the FET channel; (c) Equivalent electric circuit diagram for the graphene channel: TL1 and TL2 are transmission lines representing the top-gated and back-gated plasmonic cavities, $C_{b1}$ and $C_{b2}$ are the fringing capacitances.

FIG. 2. Complex frequencies $\omega = \omega' + i\omega''$ of the band plasmons *vs* the Bloch wave number $k$ in the density-modulated FET channel at different values of the Mach number $M = v_{02}/v_{p2}$. Here $v_{02}$ and $v_{p2}$ are the drift velocity and the plasma velocity, respectively in the back-gated cavities, $\omega_0 = v_{p2}/L$, $L$ is the crystal period. All other parameters are defined in the text.

FIG. 3. Evolution of the plasmonic spectrum at the center of the Brillouin zone ($k = 0$) with increasing electron density $n_{01}$ in the grating gated cavities. Here $n_{02} = const$ is the electron density in the back-gated cavities, $\omega_0 = v_{p2}/L$. All other parameters are defined in the text.

FIG. 4. Plasmonic spectrum at the center of the Brillouin zone ($k = 0$) as a function of the drift velocity $v_{02}/v_F$ at different values of the asymmetry factor $\gamma = C_{b2}/C_{b1}$ and direction of the drift velocity $v_0$: (a) $\gamma = 1$, $v_0 \lessgtr 0$; (b) $\gamma = 10$, $v_0 > 0$; (c) $\gamma = 10$, $v_0 < 0$.

FIG. 5. Frequencies ($f$) and increments/decrements ($\omega''$) of the plasma modes exited by an external EM wave in the graphene dual-grating-gate TeraFET as a function of the drift velocity $v_{02}$. Both the damped modes (dashed grey lines) and the active mode (red and blue lines) are shown; $f_s$ is the threshold mode frequency with the quality factor $Q = 1$; $v_{02}^{A,B}$ are the drift velocities at the boundaries (dashed vertical lines) of the gap in the detected plasmonic spectrum.

FIG. 6. The structure geometry and the static electric field distribution near the boundary between the gated and the ungated parts of the transistor channel (a) and the equivalent electric circuit describing this boundary (b).

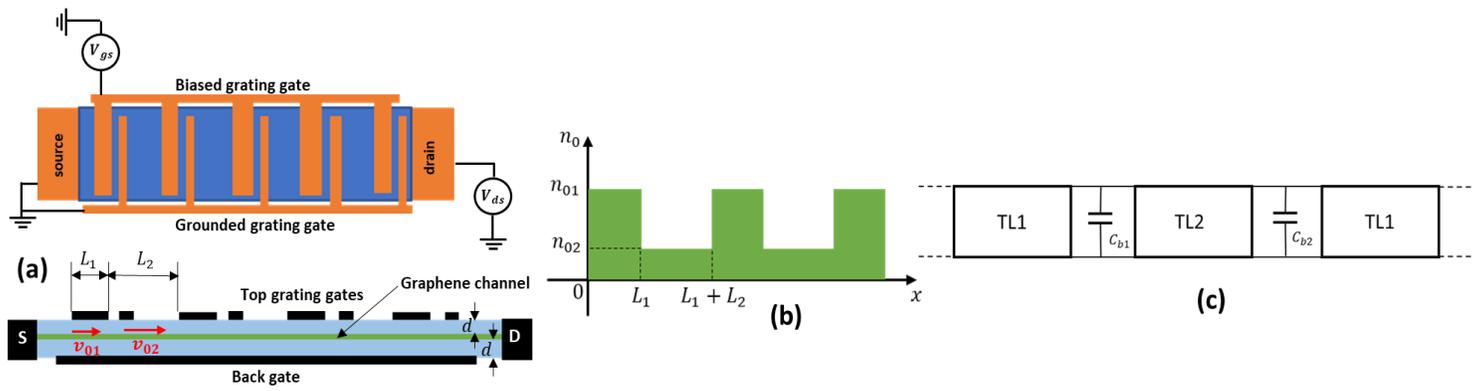

**Figure 1**

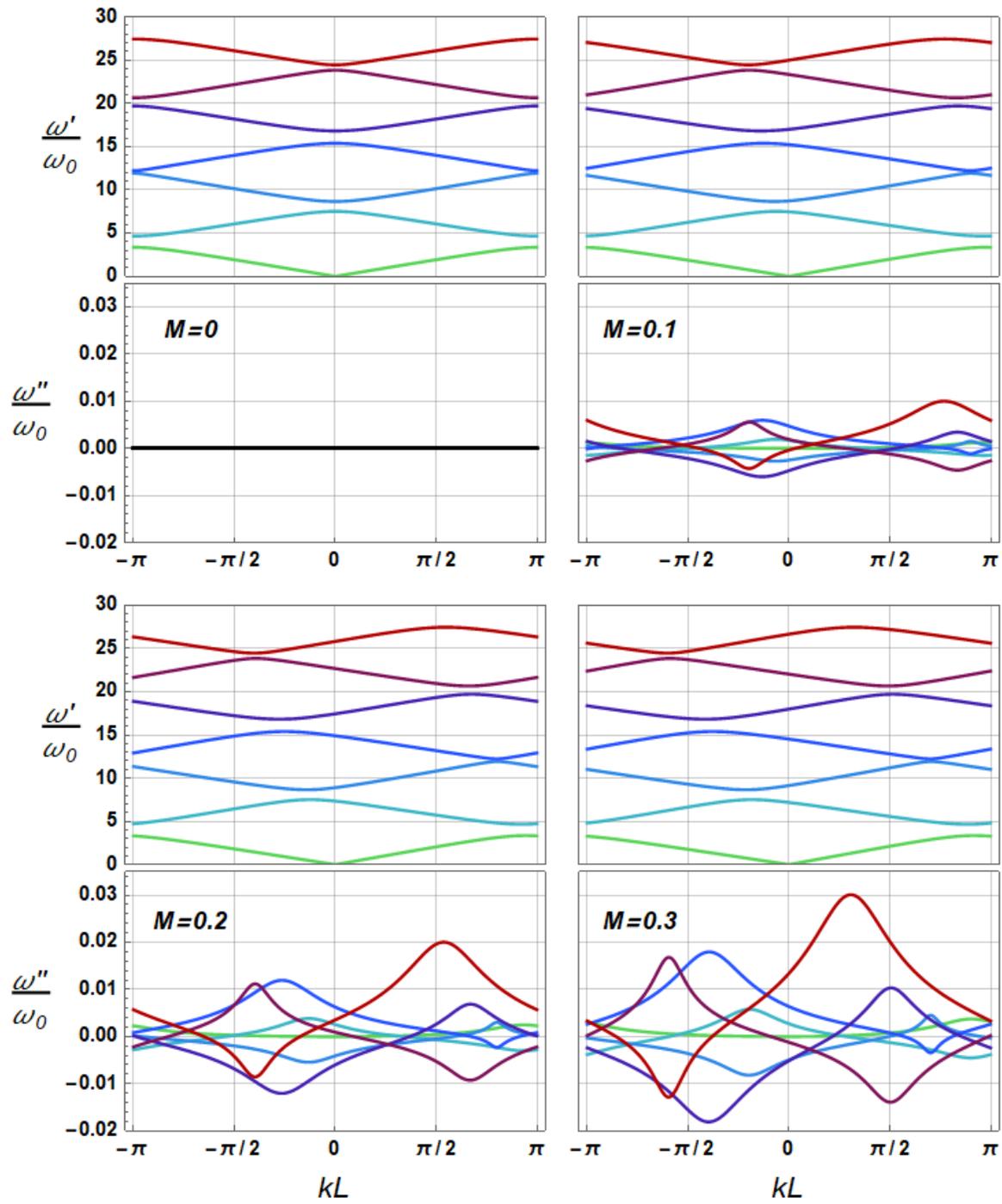

**Figure 2**

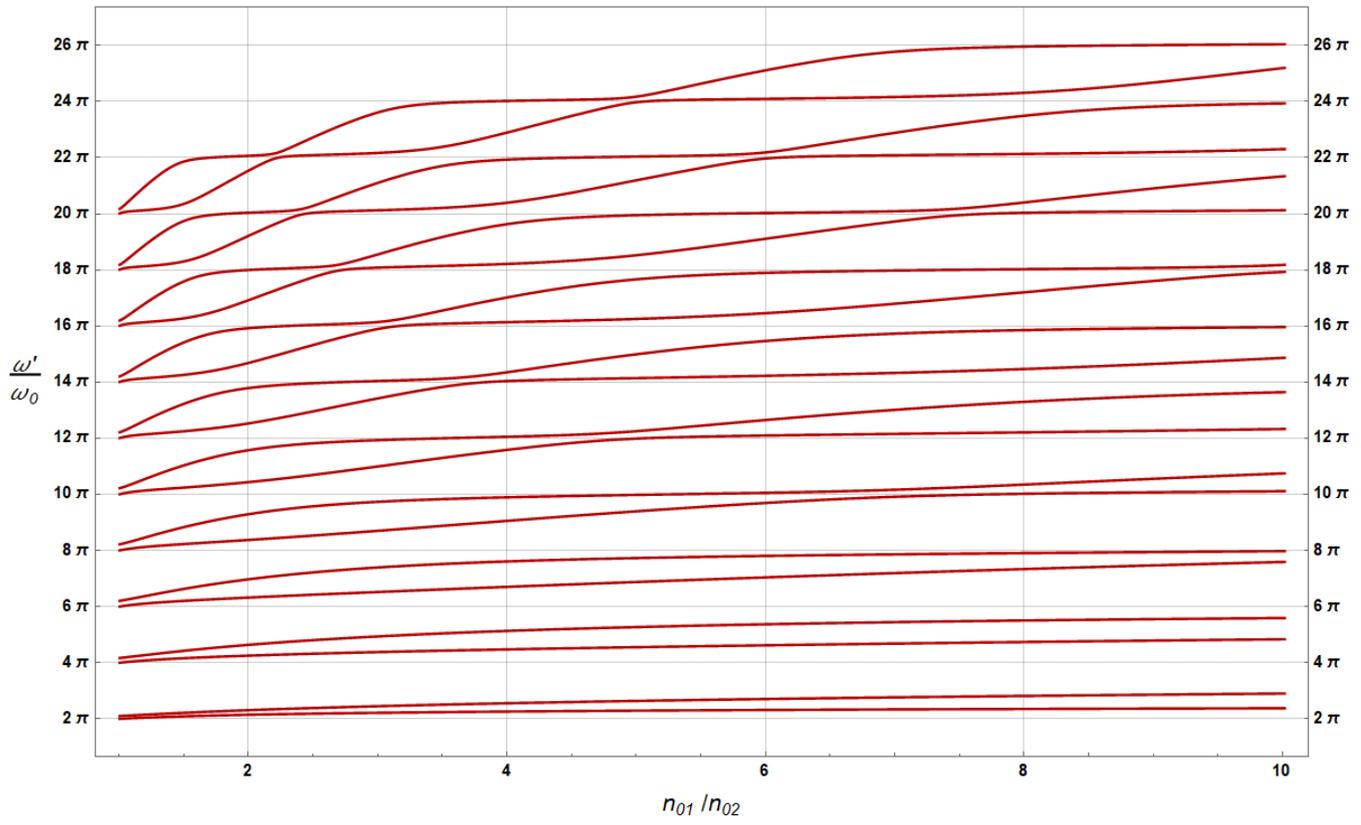

**Figure 3**

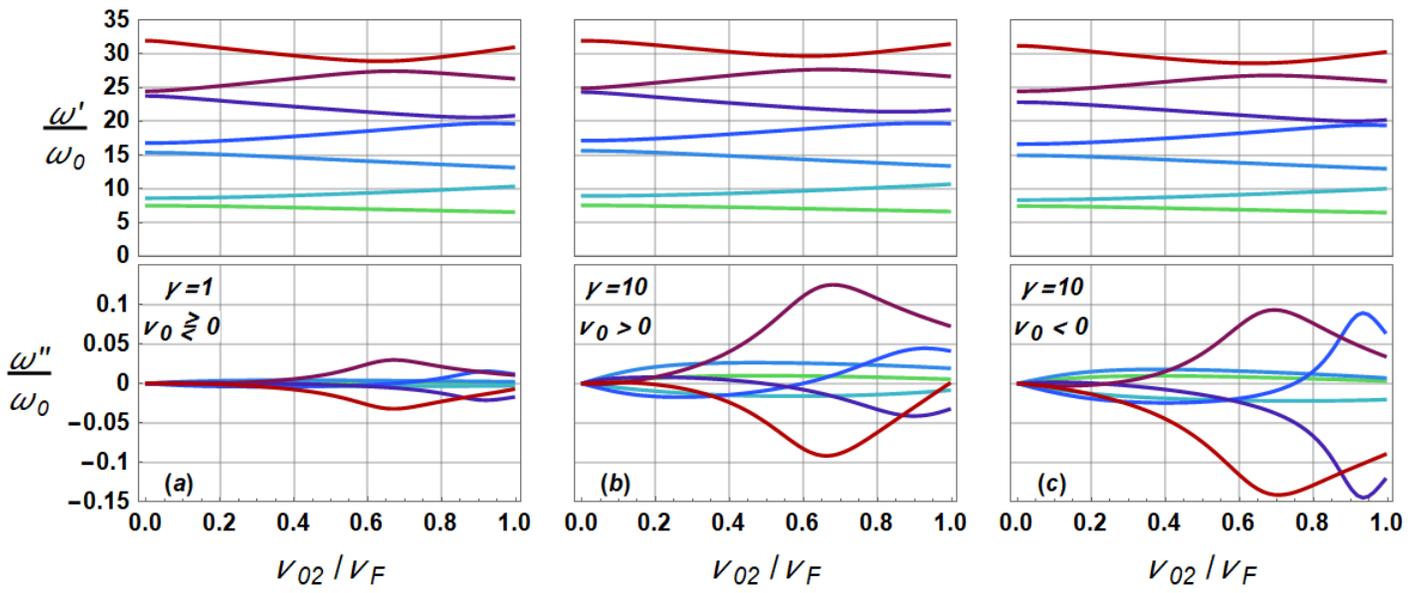

**Figure 4**

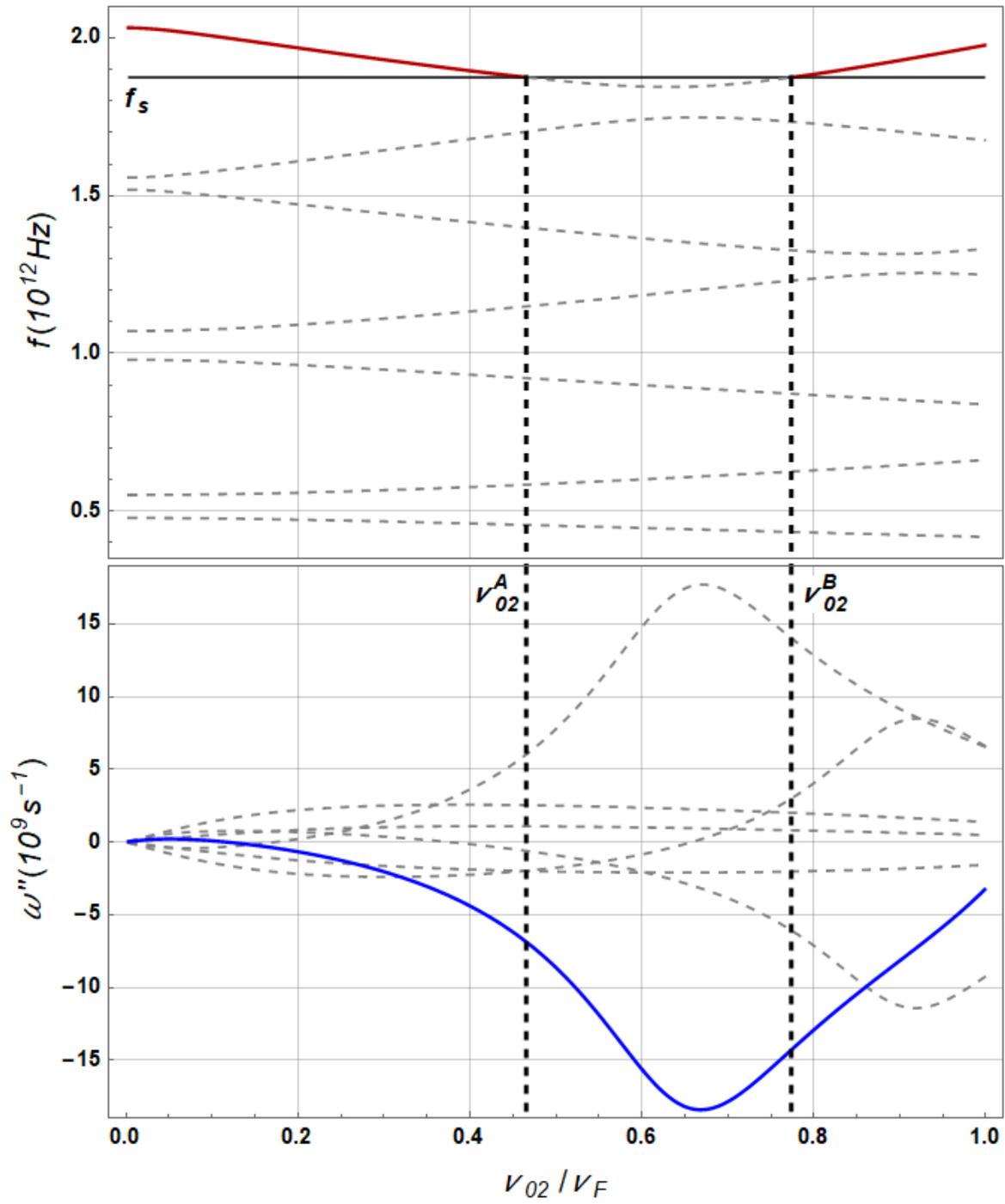

**Figure 5**

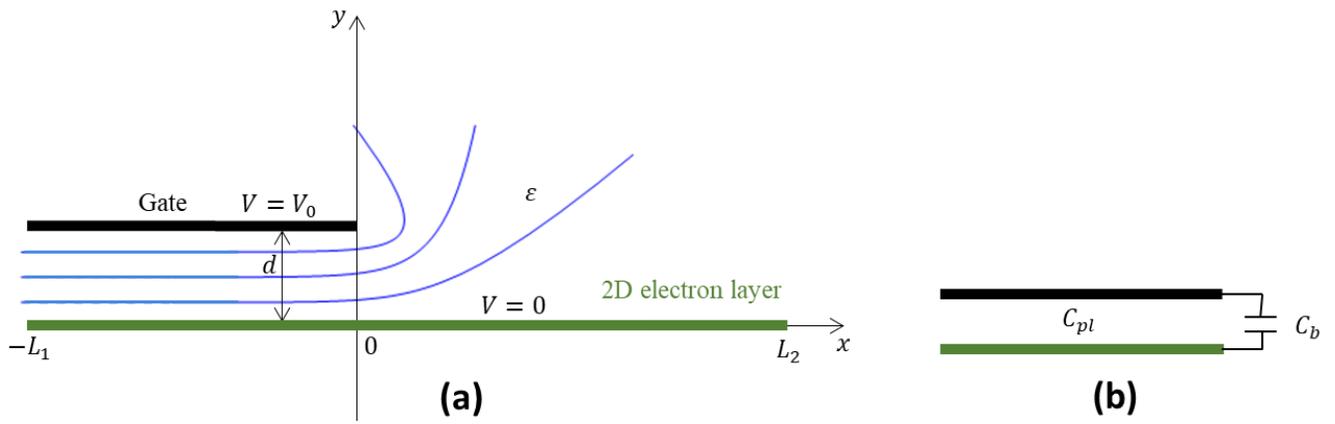

**Figure 6**